\documentclass[12pt]{article}
\usepackage[english, russian]{babel}
\usepackage[cp1251]{inputenc}
\usepackage[english]{babel}
\usepackage{amssymb,amsmath,amsthm}
\begin{document}
\begin{center}{GROUND STATES FOR POTTS MODEL WITH COMPETING INTERACTIONS\\
ON CAYLEY TREE}\end{center}
\begin{center}{BOTIROV G.I. \\
}\end{center}

\begin{center}{\textbf{Abstract}}\end{center}
We consider the Potts model with two-step interactions and spin
values 1,2,3,4 on a Cayley tree. We describe periodic ground
states and verify the Peierls condition for the model.

\textbf{Keywords:} Cayley tree; Configuration; Potts model; Ground
states; Peierls condition.

\section{Introduction}

The state of a physical system having the lowest possible
potential energy. For example, an electron in the lowest energy
orbital in a hydrogen atom is in a ground state. The ground state
of a physical system tends to be stable unless energy is applied
to it from the outside; states that are not the ground state have
a tendency to revert to the ground state, giving off energy in the
process.

The ground states for models on the cubic lattice $Z^d$ were
studied in many works (see e.g. [5,6,8,9]). The Ising model, with
two values of spin $±1$ was considered and became actively
researched in the 1990’s and afterwards (see for example
[1–3,10]).

In [11] an Ising model on a Cayley tree with competing
interactions is considered. The goal of this paper is to study of
(periodic and non periodic) ground states and to verify the
Peierls condition for the Potts model with competing interactions.

For the Ising model with competing interactions, in [12] the set
of all weakly periodic ground states corresponding to normal
divisors of indices 2 and 4 of the group representation of the
Cayley tree is described.

In the Pirogov-Sinai theory configurations can be described by
contours which satisfy Peierls condition. This theory provides
tools for a very detailed knowledge of structure of Gibbs measures
in a region in relevant parameters space [13].

Pirogov and Sinai developed a theory of phase transitions in
systems satisfying Peierls condition. W. Holsztynski and J. Slawny
give a criterium for the Peierls condition to hold and apply it to
several systems. In particular they proved that ferromagnetic
system satisfies the Peierls condition iff its (internal) symmetry
group is finite. And using an algebraic argument they show that in
two dimensions the symmetry groups of reduced translation
invariant systems is finite [7].

The Potts model is a generalization of the Ising model in which
each lattice site contains an entity (a spin) that can be in one
of $q$ states. Potts models are useful for describing the
absorption of molecules on crystalline surfaces and the behavior
of foams, for example, and exhibit a discontinuous or continuous
phase transition depending on the value of $q$.

The paper is organized as follows. In Section 2 we give
definitions of Cayley tree, configuration space and the Potts
model with competing interactions. In Section 3 we construct
periodic ground states. In Section 4 we check Peierls condition
for the model.

\section{Definitions and statement of problem}

\textbf{2.1. Cayley tree.}

The Cayley tree $\Gamma^k$ of order $k \geq 1$ is the infinite
tree (i.e., a cycle-free graph) each of whose vertices has exactly
$k + 1$ outgoing edges. Let $\Gamma^k = (V,L, i)$, where $V$ is
the set of vertices of $\Gamma^k$, and $L$ is the set of edges,
and $i$ is the incidence function, which takes each edge $l \in L$
to its endpoints $x, y \in V$. If $i(l) = \{x, y \}$, then the
vertices $x$ and $y$ are called nearest neighbors, and we write
$<x, y>$. The distance $d(x, y), x, y \in V$, on the Cayley tree
is defined by the formula $$d(x, y) =\textrm{min}\{d : \ \exists x
=x_0, \ x_1, \dots , x_{d-1}, \ x_d = y \in V \atop   \hbox{such \
that} \ x =<x_0, x_1>, \dots , <x_{d-1}, x_d> \}.$$

There exists a one-to-one correspondence between the set $V$ of
vertices of the Cayley tree of order $k \geq 1$ and the group
$G_k$ that is the free product of $k + 1$ cyclic groups of order 2
with generators
$a_1, a_2, \dots , a_{k+1}$ [8]. \\

\textbf{2.2 Configuration space.}

Let $\Phi= \{1, 2, \dots , q\}, \ q \geq 2$. A configurations on
$V$ is defined as a function $x \in V \rightarrow \sigma(x) \in
\Phi$; the set of all configurations coincides with
$\Omega=\Phi^V$. Let $A\subset V$. By $\Omega_A$ we denote the
space of configurations defined on $A$. \\

\textbf{2.3 The Potts model.}

R.B. Potts defined the former model in 1952, at the suggestion of
C. Domb. He actually defined two models. The first is known as the
"$Z_n$ model", and supposes that at each site of a lattice there
is a two-dimensional unit vector which can point in one of $N$
equally spaced directions. Two adjacent vectors interact with
interaction energy proportional to their scalar product.

The second model is the one that will be discussed here, and
referred to simply as "the Potts model". This can be formulated on
any graph $\wp$, i.e. on any set of sites, and edges joining pairs
of sites.

We consider the Potts model with competing interactions, where the
spin takes values in the set $\Phi=\{0,1,3, \dots, q\}$, on the
Cayley tree which is defined by the following Hamiltonian

$$H(\sigma)=J_1 \sum\limits_{<x,y>, \atop x,y \in V} \delta_{\sigma(x) \sigma(y)}+J_2 \sum\limits_{x,y \in V,
 \atop d(x,y)=2} \delta_{\sigma(x) \sigma(y)} \eqno (1)$$
where $J_1, J_2 \in R$ are coupling constants and $\sigma$ a
configuration on $V$ [4].

In this paper we consider the case $q=4$.

\section{Ground states}

Ground states can be defined in two ways, yielding, roughly,
zero-temperature version of Gibbs states and equilibrium states,
respectively. The first definition is that any local perturbation
of the state increases its energy; the second, applicable to
periodic configurations, is that specific energy is minimal.

For a pair of configurations $\sigma$ and $\varphi$ coinciding
almost everywhere, i.e., everywhere except at a finite number of
points, we consider the relative Hamiltonian $H(\sigma, \varphi)$,
of the difference between the energies of the configurations
$\sigma$ and $\varphi$, i.e.,
$$H(\sigma, \varphi)=J_1 \sum \limits_{<x,y> \atop x,y \in V}(\delta_{\sigma(x)\sigma(y)}-\delta_{\varphi(x)\varphi(y)})+
J_2 \sum \limits_{x,y \in V: \atop
d(x,y)=2}(\delta_{\sigma(x)\sigma(y)}-\delta_{\varphi(x)\varphi(y)})
\eqno(2)$$ where $J = (J_1, J_2) \in R^2$ is an arbitrary fixed
parameter.

We suppose that $M$ is the set of unit balls with vertices in $V$.
The restriction of a configuration $\sigma$ on a ball $b \in M$ is
called a bounded configuration $\sigma_b$. We define the energy of
the configuration $\sigma_b$ on the ball $b$ as

$$U(\sigma_b) \equiv U(\sigma_b,J)=\frac{1}{2}J_1 \sum \limits_{<x,y>, \atop x,y \in b} \delta_{\sigma(x)\sigma(y)}+
J_2\sum \limits_{x,y \in b: \atop d(x,y)=2}
\delta_{\sigma(x)\sigma(y)}  \eqno (3)$$ where $J=(J_1,J_2) \in
R$.

Using a combinatorial calculations one can prove the following

\textbf{Lemma 3.1.} 1) Let $\sigma_b$ be a configuration with
$\sigma_b(c_b)=i$,\ (where $c_b$ is the center of the ball $b$),
and  $ |\{x:\sigma_b(x)=1\}|=m, \ |\{x: \sigma(x)=2\}|=n, \ |\{x:
\sigma(x)=3\}|=l, \ |\{x : \sigma(x)=4\}|=r $. Here $|A|$ denotes
the number of elements of $A$. Then $U(\sigma_b)$ has the
following form
$$U(\sigma_b)\equiv
U_{i,k}(m,n,l,r,
J_1,J_2)=\frac{1}{2}(\delta_{1i}m+\delta_{2i}n+\delta_{3i}l+\delta_{3i}l)J_1+(C_m^2+C_n^2+C_l^2+C_r^2)J_2
\eqno(4) $$ where $m, n, l, r \in N\cup\{0\}, \ m+n+l+r=k+1$ and
$J = (J_1, J_2) \in R^2$.

2)  For any configuration $\sigma_b$ we have
$$U(\sigma_b) \in
\{U_{i,k}(m,n,l,r, J_1,J_2): \ m, n, l,r, \in N\cup\{0\}, \  {\rm
with}\atop \ m+n+l+r=k+1  \ {\rm and}  \ J = (J_1, J_2) \in
R^2\}.$$

The following lemma can be easily proved.

\textbf{Lemma 3.2.} Relative Hamiltonian (2) has the form
$$H(\sigma, \varphi)=\sum \limits_{b \in M}(U(\sigma_b)-U(\varphi_b)).$$

Denote $$F_p^{(i)} \equiv F_p^{(i)}{(\sigma_b)}=\{j \in N_k:
\sigma_b(c_b)=i, \  \sigma_b(a_j)=p\}, \ p=1,2,3,4; \atop
\Omega_{m,n,l}^{(i)}=\{\sigma_b: \ \sigma_b(c_b)=i, \
|F_1^{(i)}|=m, |F_2^{(i)}|=n, |F_3^{(i)}|=l \}. $$

Let $S_4$ be the group of permutations on $\{1,2,3,4\}$.

$C_{m,n,l}^{(i)}=\bigcup \limits_{\pi \in
S_4}\pi(\Omega_{m,n,l}^{(i)}) $, where for $\pi=(\pi(1), \pi (2),
\pi(3),\pi(4)) \in S_4$ we put $\pi (\Omega_{m,n,l}^{(i)})=\{ \pi
\sigma: \sigma \in \Omega_{m,n,l}^{(i)}\}$ with $(\pi
\sigma)(x)=\pi(\sigma(x))$.

We define a $\hat{G}_k$ -- periodic configuration as a
configuration $\sigma(x)$ that is invariant under a finite-index
subgroup $\hat{G}_k \in G_k$, i.e., $\sigma(yx)=\sigma(x)$ for any
$x \in G_k$ and $y \in \hat{G}_k$. For a given periodic
configuration, the index of the subgroup is called the
configuration period. A configuration which is invariant under all
translations is said to be translationally invariant.

\textbf{Theorem 3.3.} {\it For any class $C_{m,n,l}^{(i)}$ and for
any bounded configuration $\sigma_b \in C_{m,n,l}^{(i)}$ there
exists a periodic configuration $\varphi$ with period non
exceeding 4 such that $\varphi_{b'} \in C_{m,n,l}^{(i)}$ for any
$b' \in M$ and $\varphi_b=\sigma_b$.}

\textbf{Proof:} For arbitrary given class $C_{m,n,l}^{(i)}$ and
$\sigma_b \in C_{m,n,l}^{(i)}$ we shall construct configuration
$\varphi$ as follows. Without loss of generality we can take $b$
as the ball with the center $e \in G_k$ (here $e$ is the identity
if $G_k$), i.e. $b=\{e, a_1, \cdots, a_{k+1}\}$. Assume
$\sigma_b(e)=i$ where $i=1,2,3,4$.

Consider

$$H_0^{(i)}=\{x \in G_k: \sum \limits_{j \in N_k \setminus
(F_1^{(i)}\cup F_2^{(i)})} \omega_j(x)- { \rm even}, \sum
\limits_{j \in F_2^{(i) }\cup F_3^{(i)}} \omega_j(x)- { \rm even}
\},$$

where $\omega_j(x)$ is the number of $a_j$ in $x \in G_k$. Since
$H_0^{(i)}$ is the subgroup of index 4 in $G_k$, the quotient
group has form
$G_k/H_0^{(i)}=\{H_0^{(i)},H_1^{(i)},H_2^{(i)},H_3^{(i)}\}$ with
the co sets
$$H_1^{(i)}=\{x \in G_k: \sum \limits_{j \in N_k \setminus
(F_1^{(i)}\cup F_2^{(i)})} \omega_j(x)- { \rm even}, \sum
\limits_{j \in F_2^{(i)}\cup F_3^{(i)}} \omega_j(x)- { \rm odd}
\};$$
$$H_2^{(i)}=\{x \in G_k: \sum \limits_{j \in N_k \setminus
(F_1^{(i)}\cup F_2^{(i)})} \omega_j(x)- { \rm odd}, \sum
\limits_{j \in F_2^{(i)}\cup F_3^{(i)}} \omega_j(x)- { \rm even}
\};$$
$$H_3^{(i)}=\{x \in G_k: \sum \limits_{j \in N_k \setminus
(F_1^{(i)}\cup F_2^{(i)})} \omega_j(x)- { \rm odd}, \sum
\limits_{j \in F_2^{(i)}\cup F_3^{(i)}} \omega_j(x)- { \rm odd }
\}.$$

We continue the bounded configuration $\sigma_b \in
C_{m,n,l}^{(i)}$ to whole lattice $\Gamma^k$ (which we denote by
$\varphi$) by
$$\varphi(x)=\left\{%
\begin{array}{ll}
    1, & \hbox{if $x \in H_0^{(i)}$;} \\
    2, & \hbox{if $x \in H_1^{(i)}$;} \\
    3, & \hbox{if $x \in H_2^{(i)}$;} \\
    4, & \hbox{if $x \in H_3^{(i)}$.} \\
\end{array}%
\right.$$

So we obtain a periodic configuration $\varphi$ with period 4
(=index of the subgroup); then by the construction
$\varphi_b=\sigma_b$. Now we shall prove that all restrictions
$\varphi_{b'}, \ b' \in M$ of the configuration $\varphi$ belong
$C_{m,n,l}^{(i)}$. Let $q_j(x)=|S_1(x) \cap H_j|, \ j=0,1,2,3$;
where $S_1(x)=\{y \in G_k:  <x,y> \}$, the set of all nearest
neighbors of $x \in G_k$.

Denote $Q(x)=(q_0(x),q_1(x),q_2(x),q_3(x))$. We note [15] that for
every $x \in G_k$ there is a permutation $\pi_x$ of the coordinate
of the vector $Q(e)$ (where $e$ as before is the identity of
$G_k$) such that $$\pi_xQ(e)=Q(x).$$

Moreover, it is easy to see that
$$Q(x)=\left\{%
\begin{array}{ll}
    Q(e), & \hbox{if  $ x \in H_0^{(0)}$;} \\
    (q_1(e),q_0(e),q_3(e),q_2(e)), & \hbox{if $x \in H_1^{(i)}$;} \\
    (q_2(e),q_3(e),q_0(e),q_1(e)), & \hbox{if $x \in H_2^{(i)}$;} \\
    (q_3(e),q_2(e),q_1(e),q_0(e)), & \hbox{if $ x \in H_3^{(i)}$.} \\
\end{array}%
\right.$$

Thus for any $b' \in M$ we have (i) if $c_{b'} \in H_0^{(i)}$
(where as before $c_{b'}$ is the center of $b'$) then
$\varphi_{b'}=\sigma_b$ up to rotation; (ii) if $c_{b'} \in
H_1^{(i)}$ then $\varphi_{b'}=\pi_1(\sigma_b)$ where $\pi_1=\left(%
\begin{array}{cccc}
  0 & 1 & 2 & 3 \\
  1 & 0 & 3 & 2 \\
\end{array}%
\right)$, then $\varphi_{b'}=\sigma_b$ up to rotation; (iii) if
$c_{b'} \in
H_2^{(i)}$ then $\varphi_{b'}=\pi_2(\sigma_b)$ where $\pi_2=\left(%
\begin{array}{cccc}
  0 & 1 & 2 & 3 \\
  2 & 3 & 0 & 1 \\
\end{array}%
\right)$, then $\varphi_{b'}=\sigma_b$ up to rotation; (iv) if
$c_{b'} \in
H_3^{(i)}$ then $\varphi_{b'}=\pi_3(\sigma_b)$ where $\pi_3=\left(%
\begin{array}{cccc}
  0 & 1 & 2 & 3 \\
  3 & 2 & 1 & 0 \\
\end{array}%
\right)$, then $\varphi_{b'}=\sigma_b$ up to rotation. Since the
energy of a configuration $\sigma_b$ and $\pi(\sigma_b)$ for
arbitrary $\pi \in S_4$ is equal we get $\varphi_{b'} \in
C_{m,n,l}^{(i)}$ for any $b' \in M$. Theorem is proved.

\textbf{Definition 3.4.} A configuration $\sigma$ is called a
\emph{ground state} of the relative Hamiltonian $H$ if
$$U(\sigma_b) = \min \{U_{i,k}(m,n,l, J_1,J_2):  \ i=1,2,3; \ m,n, l \in N \cup\{0\}
\atop {\rm with} \ 0 \leq m+n+l \leq k+1\} \  \hbox{for \ any} \ b
\in M.$$

Denote by $A_{i,k}(m,n,l)$ the set of points $(J_1,J_2)$ such that
$$A_{i,k}(m,n,l)=\{(J_1,J_2) : U_{i,k}(m,n,l,J_1,J_2) \leq U_{j,k}(m',n',l',J_1,J_2),
\atop {\rm for \ any} \  m', \ n' \ l' \in N\cup\{0\}, \
j=1,2,3,4\}.$$

Let $GS(H)$ be the set of all ground states of the relative
Hamiltonian $H$ (see (2)).

\textbf{Theorem 3.5.} (i) If $J=(0,0)$ then $GS(H)=\Omega$.

(ii) If $J \in A_{i,k}(m,n,l)$ then
$$GS(H)=\{\pi(\sigma_{m,n,l}^{(i)}):  \ \pi \in S_4\},$$
where $\sigma_{m,n,l}^{(i)} \in \Omega$ such that
$(\sigma_{m,n,l}^{(i)})_b \in \Omega_{m,n,l}^{(i)}$ for any $b \in
M, \ 0\leq m+n+l \leq k+1, \ i=1,2,3,4$.

\textbf{Proof:} Case (i) is trivial. (ii) for a given
configuration $\sigma_b$ which makes $U(\sigma_b)$ minimal, by
Theorem 3.3 one can construct the periodic ground states
$\sigma_{m,n,l}^{(i)}$ with period non exceeding four. For each
case the exact number of such ground states coincides with number
of the configurations $\sigma_b$ which make $U(\sigma_b)$ minimal.
Theorem is proved.

\section{The Peierls condition}

Denote by $\textbf{U}$ the collection of all possible values of
$U(\sigma_b)$ for any configuration $\sigma_b$, $b \in M$.

Put $U^{\textrm{min}}=\textrm{min}\{U:U \in \textbf{U}\}$ and
$$\lambda_0=\textrm{min}\{\textbf{U} \backslash \{U \in
\textbf{U}: U=U^{\textrm{min}}\}\}-U^{\textrm{min}}.$$
\\ \textbf{Definition 4.1.} Let $GS$ be the set of
all ground states of relative Hamiltonian $H$. A ball $b \in M$ is
said to be an \emph{improper} ball of the configuration $\sigma$
if $\sigma_{b} \neq \varphi_b$ for any $\varphi \in GS$. The union
of improper balls of a configuration $\sigma$ is called the
\emph{boundary} of the configuration and denoted by $\partial(\sigma)$.\\
\textbf{Definition 4.2.} The relative Hamiltonian $H$ with the set
of ground states $GS$ satisfies the Peierls condition if for any
$\varphi \in GS$ and any configuration $\sigma$ coinciding almost
everywhere with $\varphi$ (i.e. $|\{x \in V: \sigma(x) \neq
\varphi(x)\}| < \infty$) $$H(\sigma, \varphi) \geq \lambda
|\partial(\sigma)|$$ where $\lambda $ is a positive constant which
does not depend on $\sigma$, and $|\partial(\sigma)|$ is the
number of balls in $\partial(\sigma)$.

\textbf{Theorem 4.3} If $J \neq (0,0)$ then the Peierls condition
is satisfied.

\textbf{Proof:} Note that $|\textbf{U}|=1$ if and only if $J
=(0,0)$, consequently $\lambda_0
>0$ if $J\neq (0,0)$.

Suppose $\sigma$ coincides almost everywhere with a ground state
$\varphi \in GS(H)$ then we have $U(\sigma_b)-U(\varphi_b) \geq
\lambda_0$ for any $b \in \partial(\sigma)$ since $\varphi$ is a
ground state. Thus
$$H(\sigma,\varphi) = \sum \limits_{b \in M}(U(\sigma_b)-U(\varphi_b))= \sum \limits_{b \in \partial(\sigma)}(U(\sigma_b)
-U(\varphi_b)) \geq \lambda_0|\partial(\sigma)|.$$

Therefore, the Peierls condition is satisfied for
$\lambda=\lambda_0$. The theorem is proved.

\textbf{Acknowledgments. } The work supported by Grants $\Phi A
\Phi 1-\Phi 09$ and  $\Phi A \Phi 1 - \Phi 003 + \Phi 067$  of CST
of the Republic Uzbekistan.  The author especially express their
thanks to Professor U.A. Rozikov  for the useful discussions.

\begin{center} REFERENCES \end{center}

1. Baxter R.J., {\it Exactly  Solved Models in Statistical
Mechanics}, Academic Press, London, New York, 1982.

2. Bleher P.M., Ruiz J. and Zagrebnov V.A., On the purity of the
limiting Gibbs state for the Ising model on the Bethe lattice,
{\it Jour. Statist. Phys.} 79:473–482 (1995).

3. Bleher P.M., Ruiz J., Schonmann R. H., Shlosman S. and
Zagrebnov V.A., Rigidity of the critical phases on a Cayley tree,
{\it Moscow Math.Journ.} 3:345–362 (2001).

4. Botirov G.I., Rozikov U.A., Potts model with competing
interactions on the Cayley tree: the contour method, {\it
Theoretical and Mathematical Physics}, 153(1): 1423–1433 (2007)

5. Ganikhodjaev N.N., Rozikov U.A., A description of periodic
extremal Gibbs measures of some lattice models on the Cayley tree,
{\it Theor. Math. Phys.} {\bf 111}: 480-486 (1997).

6. Georgii H.O., Gibbs measures and phase transitions (Walter de
Gruyter, Berlin, 1988.

7. Holsztynski W. and J.Slawny, Peierls condition and the number
of ground states. {\it Commun.Math. Phys}. 61:177–190 (1978).

8. Pirogov S.A., and Sinai Y.G., Phase diagrams of classical
lattice systems, I. {\it Theor. Math. Phys.} 25:1185–1192 (1975).

9. Pirogov S.A., and Sinai Y.G., Phase diagrams of classical
lattice systems, II. {\it Theor. Math. Phys.} 26:39–49 (1976).

10. Rozikov U.A., Partition structures of the Cayley  tree and
applications for describing periodic Gibbs distributions. {\it
Theor. Math. Phys.} {\bf 112}: 929--933 (1997).

11. Rozikov U.A., A constructive description of ground states and
Gibbs measures for Ising model with two-step interactions on
Cayley  tree. {\it Journal of Statistical Physics Phys.} {\bf
122}: 217--235 (2006).

12. Rozikov U.A. and Rakhmatullaev, Description of weakly periodic
Gibbs measures for the Ising model on a Cayley tree, {\it
Theoretical and Mathematical Physics,} 156(2): 1218–1227 (2008).

13. Sinai Ya.G., {\it Theory of phase transitions: Rigorous
Results} (Pergamon, Oxford, 1982).

Institute of mathematics and information technologies, Tashkent,
UZBEKISTAN \\ E-mail: botirovg@yandex.ru

\end{document}